\documentclass[journal]{IEEEtran}

\usepackage{url}  

\usepackage{multirow}
\usepackage{graphicx}
\usepackage{amsmath,amssymb,bm}

\usepackage{xcolor}
\usepackage[colorlinks=true,citecolor=blue,urlcolor=blue]{hyperref}
\usepackage{cite}

\usepackage{booktabs}
\usepackage{array}

\usepackage{fancyhdr}

\begin{document}

	\title{Unsupervised Bidirectional Cross-Modality Adaptation via Deeply Synergistic Image and Feature Alignment for Medical Image Segmentation}

	\author{
		Cheng~Chen,~\IEEEmembership{Student Member,~IEEE,}
		Qi~Dou,~\IEEEmembership{Member,~IEEE,}\\
		Hao~Chen,~\IEEEmembership{Member,~IEEE,}
		Jing~Qin,~\IEEEmembership{Member,~IEEE,}
		Pheng~Ann~Heng,~\IEEEmembership{Senior Member,~IEEE}
		\thanks{Manuscript received July 25, 2019; revised December 23, 2019; accepted February 3, 2020. This work was partially supported by HK RGC TRS project T42-409/18-R, Hong Kong Innovation and Technology Fund project No. ITS/426/17FP, project No. ITS/311/18FP and CUHK T Stone Robotics Institute, and a grant from the Hong Kong Research Grants Council (No. PolyU 152035/17E). (Corresponding author: Qi Dou.)}%
		\thanks{C. Chen and Q. Dou are with the Department of Computer Science and Engineering, The Chinese University of Hong Kong, Hong Kong, China (e-mail: cchen, qdou@cse.cuhk.edu.hk).}%
		\thanks{H. Chen is with the Department of Computer Science and Engineering, The Chinese University of Hong Kong, Hong Kong, China, and the Imsight Medical Technology Co. Ltd., China (e-mail: hchen@cse.cuhk.edu.hk).}%
		\thanks{J. Qin is with the Centre for Smart Health, School of Nursing, The Hong Kong Polytechnic University, Hong Kong, China (e-mail: harry.qin@polyu.edu.hk).}%
		\thanks{P. A. Heng is with the Department of Computer Science and Engineering, and the T Stone Robotics Institute, The Chinese University of Hong Kong, Hong Kong, China (e-mail: pheng@cse.cuhk.edu.hk).}%
	}

	\markboth{IEEE Transactions on Medical Imgaging}
	{Shell \MakeLowercase{\textit{et al.}}: Bare Demo of IEEEtran.cls for Journals}
		
	\maketitle
	
	
	\begin{abstract}
		Unsupervised domain adaptation has increasingly gained interest in medical image computing, aiming to tackle the performance degradation of deep neural networks when being deployed to unseen data with heterogeneous characteristics. In this work, we present a novel unsupervised domain adaptation framework, named as~\textit{Synergistic Image and Feature Alignment (SIFA)}, to effectively adapt a segmentation network to an unlabeled target domain.
		Our proposed SIFA conducts synergistic alignment of domains from both image and feature perspectives.
		In particular, we simultaneously transform the appearance of images across domains and enhance domain-invariance of the extracted features by leveraging adversarial learning in multiple aspects and with a deeply supervised mechanism. 
		The feature encoder is shared between both adaptive perspectives to leverage their mutual benefits via end-to-end learning.
		We have extensively evaluated our method with cardiac substructure segmentation and abdominal multi-organ segmentation for bidirectional cross-modality adaptation between MRI and CT images.
		Experimental results on two different tasks demonstrate that our SIFA method is effective in improving segmentation performance on unlabeled target images, and outperforms the state-of-the-art domain adaptation approaches by a large margin.
	\end{abstract}
	
	
	\section{Introduction}
	\begin{figure}
		\centering
		\includegraphics[width=0.49\textwidth]{{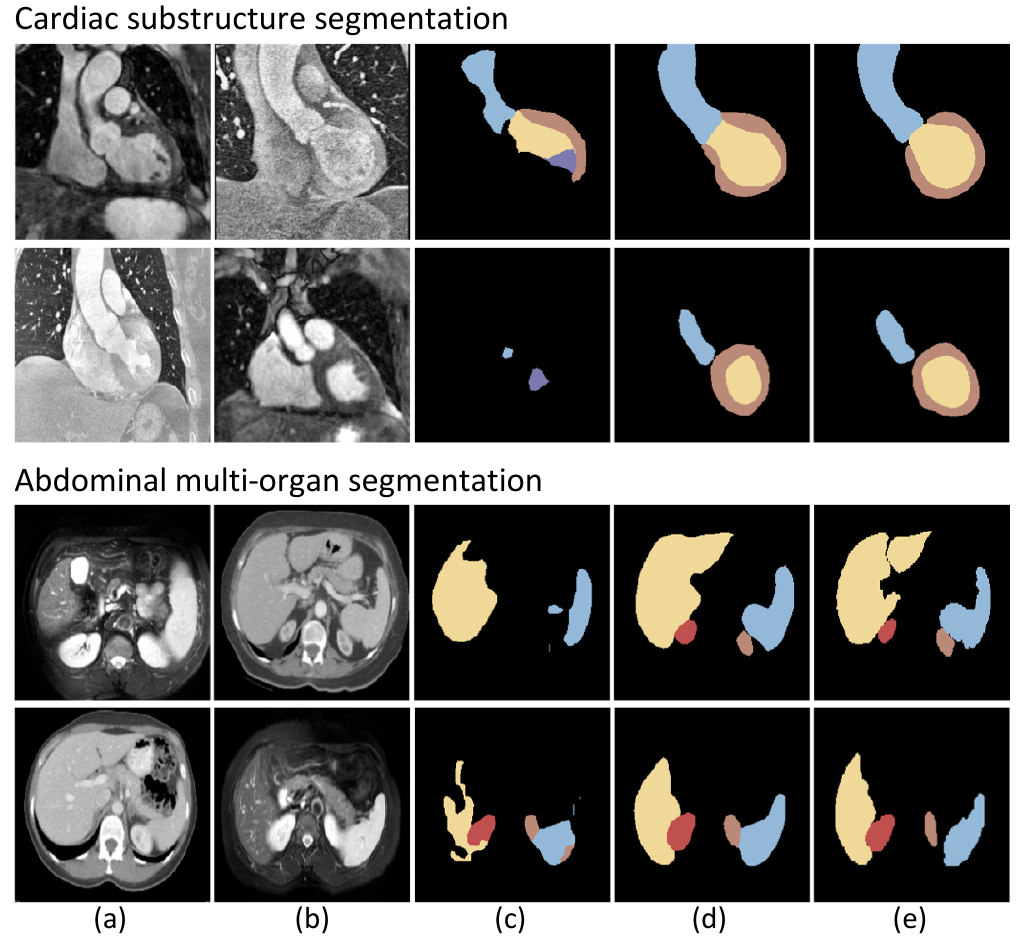}}
		\caption{Illustration of severe cross-modality domain shift and effectiveness of our method on cardiac images (top) and abdominal images (bottom). The first and third rows show MRI to CT adaptation, and the second and fourth rows show CT to MRI adaptation: a) examples of training images, b) examples of testing images, c) segmentation results of testing images without adaptation, d) segmentation results using our proposed SIFA, e) the ground truths.}
		\label{fig:intro}
	\end{figure}
	
	\IEEEPARstart{D}{eep} neural networks have achieved great success, when a large amount of labelled data are available and the training and testing data are drawn from the same distribution~\cite{ronneberger2015u,cciccek20163d,milletari2016v}. 
	However, well-trained models often fail when deployed to real-world clinical scenarios, as medical images acquired with different acquisition parameters or modalities have very different characteristics~\cite{ghafoorian2017transfer,gibson2018inter,kamnitsas2017unsupervised,DBLP:conf/ijcai/DouOCCH18}.
	For instance, in Fig.~\ref{fig:intro}, the cardiac area and abdominal area can be imaged by both magnetic resonance imaging (MRI) and computed tomography (CT), but with significantly different visual appearance given the different physical principles of imaging modalities.
	Such cross-modality domain shift would lead to severe performance degradation of deep networks. 
	Although it is not difficult for human eyes to recognize the same anatomy across modalities, the deep neural networks trained on MRI data may completely fail in segmenting CT images and vice versa.  
	
	To reduce the performance degradation caused by domain shift, research works on domain adaptation of deep models are emerging, aiming to effectively transfer the knowledge learned from the source domain to the target domain. 
	A straightforward way is to fine-tune models pre-trained on the source data with additional labeled target data~\cite{ghafoorian2017transfer}. However, annotating data for each new domain is prohibitively expensive or sometimes even infeasible, especially in medical areas requiring expert knowledge.
	Instead, unsupervised domain adaptation is practically more appealing, for which no labels in the target domain are required.
	
	Existing methods on unsupervised domain adaptation generally involve aligning the source and target distributions from two perspectives. 
	One stream is the \textit{image alignment}, which transforms the image appearance between domains with an image-to-image transformation model~\cite{DBLP:conf/iccv/ZhuPIE17,russo2017source,zhang2018task,bousmalis2017unsupervised,zhao2018supervised}. 
	The other stream focuses on \textit{feature alignment}, which aims to extract domain-invariant features usually by minimizing feature distance between domains via adversarial learning~\cite{ganin2016domain,tzeng2017adversarial,DBLP:conf/ijcai/DouOCCH18,tsai2018learning,sankaranarayanan2018learning,wang2019patch}.
	We recognize that the image alignment and feature alignment address domain shift from complementary perspectives, i.e., \emph{image alignment} at the input level and \emph{feature alignment} at the feature level of a deep neural network. 
	Thus, we consider that performing both alignment aspects in one unified framework can leverage their individual advantages to improve domain adaptation performance.  
	Importantly, the combination between image and feature alignment should be a synergistic merge to exploit their mutual interactions and benefits, which has not been tapped in previous works~\cite{hoffman2017cycada,zhang2018fully}.
	
	In this work, we present a novel unsupervised domain adaptation framework, called \textit{Synergistic Image and Feature Alignment (SIFA)}, which introduces synergistic fusion of alignments from both image and feature perspectives.
	Specifically, we transform the labeled source images to the appearance of target data, using generative adversarial networks with a cycle-consistency loss.
	When using the synthesized target-like images to train a segmentation model, we further integrate the feature alignment to combat the remaining domain shift by using adversarial learning in multiple aspects and with a deeply supervised mechanism.
	Here, for more effective feature alignment, we connect discriminators to the semantic segmentation predictions and source-like images, which are generated from the encoded features of the synthesized or real target images. 
	Importantly, in our designed SIFA framework, the feature encoder is shared, 
	such that the entire domain adaptation framework is unified and both image and feature alignments are seamlessly integrated into an end-to-end learning framework. 
	We apply our method to the challenging cross-modality adaptation with severe domain shift to validate its effectiveness.
	Our main contributions are summarized as: 
	
	\begin{itemize}
		
		\item We investigate the important yet challenging problem of unsupervised domain adaptation for medical image segmentation. 
		We present a novel framework, SIFA, which exploits synergistic image and feature alignments to address domain shift from complementary perspectives.
		
		\item We enhance the feature alignment by applying adversarial learning in two aspects, i.e., semantic prediction space and generated image space, and incorporating the deeply supervised mechanism on top of the adversarial learning.
		
		\item We conduct extensive experiments of bidirectional cross-modality adaptation between MRI and CT on two multi-class segmentation tasks, i.e., cardiac substructure segmentation and abdominal multi-organ segmentation.
		Our method significantly improves segmentation performance on unlabeled target images, and outperforms the state-of-the-art approaches.
		Notably, in the abdominal dataset, our unsupervised domain adaptation achieves results which are very close to the supervised training upper bound. 
		The code is available at \url{https://github.com/cchen-cc/SIFA}.
		
	\end{itemize}
	
	This work is a significant extension of our prior conference paper~\cite{chen2019synergistic}, regarding the following highlighted points.
	First, we further improve our method by incorporating deeply supervised feature alignment to increase adaptation performance. 
	Second, we explore adaptation in bidirections, i.e., both MRI to CT and CT to MRI.
	Third, we enhance the experimental validation by adding a new task of abdominal multi-organ segmentation with very impressive results.
	Fourth, we conduct comprehensive comparison with more state-of-the-art methods to demonstrate the effectiveness of SIFA.
	In addition, we comprehensively discuss the experimental results and the strengths and limitations of the proposed method. A more thorough literature review of related works is also included.

	\section{Related Work}
	Domain shift has been a long-standing problem in medical image analysis due to the common inter-scanner or cross-modality variations~\cite{wang1998correction,nyul1999standardizing,heimann2013learning,bermudez2016scalable}. 
	Deep domain adaptation has recently been an active research field to transfer knowledge learned from the source domain to the target data either in a supervised or unsupervised manner. 
	In \cite{ghafoorian2017transfer} and \cite{van2015transfer}, supervised transfer learning is employed to reduce the required amount of annotations in the target domain for segmentation task across MRI datasets. 
	In \cite{karani2018lifelong}, the batch normalization layers of the source model are adapted with labeled target MRI images in a lifelong learning setting. 
	However, these methods require additional labeled target data. Instead, unsupervised domain adaptation with zero extra target domain labels is more desirable. 
	Plenty of adaptive approaches have been proposed with different strategies~\cite{bousmalis2017unsupervised,zhao2018supervised,ganin2016domain,DBLP:conf/icml/LongC0J15,sun2016deep,perone2019unsupervised,van2018learning,dou2019domain}.
	In this section, we focus on adversarial learning based unsupervised domain adaptation since it is highly related to our work.
	
	Recent advances adopt adversarial learning to address domain shift from different perspectives, including the image-level alignment, feature-level alignment and their mixtures.
	With a gratitude to generative adversarial network~\cite{DBLP:conf/nips/GoodfellowPMXWOCB14},
	image alignment methods have been developed to tackle domain shift at the input level to networks, by transforming the source images to appear like the target ones or vice versa~\cite{bousmalis2017unsupervised,shrivastava2017learning,hoffman2017cycada,russo2017source,zhao2018supervised,yang2018generalizing}. 
	With the wide success of CycleGAN~\cite{DBLP:conf/iccv/ZhuPIE17} in unpaired image-to-image transformation, many previous image alignment approaches are based on the CycleGAN framework with additional constrains to further regularize the image transformation process.
	For example, both \cite{zhang2018task} and \cite{chen2018semantic} introduce semantic consistency into the CycleGAN to facilitate the transformation of target X-ray images towards the source images for testing with the pre-trained source models. 
	For cross-modality adaptation, Jiang et al.~\cite{jiang2018tumor} first transform CT images to resemble MRI appearance using CycleGAN with tumor-aware loss, then the generated MRI images are combined with a few real MRI data for semi-supervised tumor segmentation.
	In \cite{huo2018synseg} and \cite{zhang2018translating}, CycleGAN is combined with a segmentation network to compose an end-to-end framework. 
	Compared to \cite{huo2018synseg} and \cite{zhang2018translating}, a key characteristic of our approach is the shared encoder for both image transformation and segmentation task. Through the parameter sharing, the image alignment and feature alignment in our framework are able to work synergistically on reducing domain shift during the end-to-end training.

	Meanwhile, another stream of works focus on feature alignment, aiming to extract domain-invariant features of deep neural networks in an adversarial learning scenario.
	Pioneering works try to apply a discriminator directly in the feature space to differentiate the features across domains~\cite{ganin2016domain,tzeng2017adversarial}. 
	Recent studies propose to project the high-dimensional feature space to other compact spaces, such as the semantic prediction space~\cite{tsai2018learning} or the image space~\cite{sankaranarayanan2018learning}, and a discriminator operates in the compact spaces to derive adversarial losses for more effective feature alignment.
	For medical applications, the cross-protocol MRI segmentation in \cite{kamnitsas2017unsupervised} makes the earliest attempts of aligning feature distributions with an adversarial loss. 
	Later on, the adversarial training is combined with other regularization to achieve better adaptation performance across ultrasound datasets~\cite{degel2018domain}, histopathology images~\cite{ren2018adversarial}, and cardiac MRI~\cite{zhang2018multi}. 
	In \cite{wang2019patch,joyce2018deep,dong2018unsupervised}, the discriminator is adopted to differentiate the semantic predictions generated from the extracted features, demonstrating more effective feature alignment.
	For cross-modality cardiac segmentation in \cite{DBLP:conf/ijcai/DouOCCH18,dou2018pnp}, the adversarial learning is employed to only adapt the early-layer feature distributions while the higher-layer features are fixed. However, their method needs comprehensive empirical studies to determine the optimal adaptation depth. By contrast, in our framework, the discriminators are applied in two compact spaces, i.e. the semantic prediction space and generated image space, without the need of specifying the intermediary layer that the discriminator is connected to.

	The image and feature alignment address domain shift from different perspectives to deep neural networks, which are in fact complementary to each other. 
	Combining these two adaptive strategies to achieve a stronger domain adaption technique is under explorable progress.
	As the state-of-the-art methods for semantic segmentation adaptation methods,
	CyCADA~\cite{hoffman2017cycada} and Zhang et al.~\cite{zhang2018fully} achieved leading performance in adaptation between synthetic to real world driving scene domains. However, their image and feature alignments are sequentially connected and trained in separate stages without interactions.
	
	Due to the severe domain shift in cross-modality medical images, feature alignment or image alignment alone may not be sufficient in this challenging task while the simultaneous alignments from the two perspectives have not been fully explored yet. 
	To tackle the challenging cross-modality adaptation for segmentation task, we propose to synergistically merge the two adaptive processes in a unified network to fully exploit their mutual benefits towards unsupervised domain adaptation.

	\section{Methods}
	\begin{figure*}
		\centering
		\includegraphics[width=0.92\textwidth]{{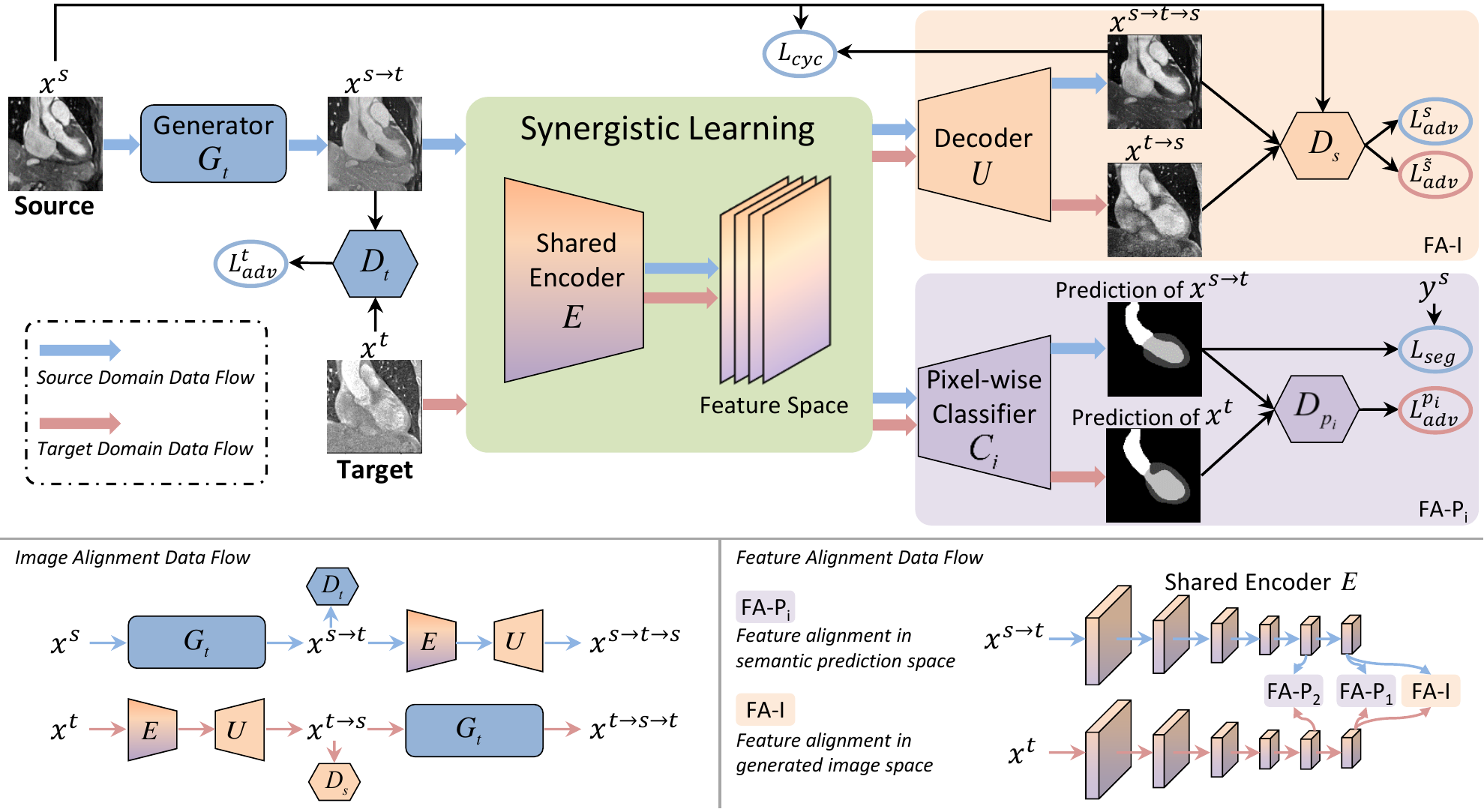}}
		\caption{Overview of our unsupervised domain adaptation framework (top), and separate views of data flows for image alignment (bottom left) and feature alignment (bottom right).
			The generator $G_t$ serves the source-to-target image transformation. The encoder $E$ and decoder $U$ form the reverse transformation, where the encoder $E$ is also connected with pixel-wise classifiers $C_i$ for image segmentation. The discriminators $\{D_t,D_s,D_{p_i}\}$ differentiate their inputs accordingly to derive adversarial losses. 
			The blue and red arrows indicate the data flows for the source and target images respectively. The target domain cycle-consistency is shown in the image alignment data flow view (bottom left) while omitted in the overview (top) for ease of illustration. Best viewed in color.}
		\label{fig:overview}
	\end{figure*}
	
	Fig.~\ref{fig:overview} presents an overview of our proposed method for unsupervised domain adaptation in medical image segmentation. 
	We propose synergistic image and feature alignment to effectively narrow the performance gap caused by domain shift.
	The two perspectives of alignments are seamlessly integrated into a unified model, and hence, both adaptive aspects can mutually benefit each other via the end-to-end training procedure.
	
	\subsection{Appearance Transformation for Image Alignment}
	Due to domain shift, images across domains usually present different visual appearance. The goal of image alignment is to narrow the domain shift by transforming the image appearance between the source and target domains. 
	Formally, given a labeled dataset $\{x_i^s,y_i^s\}_{i=1}^N$ from the source domain, and an unlabeled dataset $\{x_j^t\}_{j=1}^M$ from the target domain, we aim to transform the source images $x^s$ towards the appearance of target ones $x^t$.
	The obtained transformed images should look as if drawn from the target domain, while the original contents with structural semantics remain unaffected. 
	
	\textit{1) Appearance Transformation:} we employ generative adversarial networks for unpaired image-to-image transformation by constructing a generator $G_t$ and a discriminator $D_t$.
	The generator aims to transform the source images to target-like ones $G_t(x^s)\! = \! x^{s\to t}$.
	The discriminator competes with the generator to correctly differentiate the fake transformed image $x^{s\to t}$ and the real target image $x^t$.
	Therefore, in the target domain, the $G_t$ and $D_t$ form a minimax two-player game and are optimized via the adversarial learning:
	\begin{equation}
	\small
	\begin{split}
	\mathcal{L}_{\textit{adv}}^t(G_t,D_t) = &~\mathbb{E}_{x^{t}\sim X^{t}}[\text{log}D_t(x^{t})]+\\
	&~\mathbb{E}_{x^{s}\sim X^{s}}[\text{log}(1-D_t(G_t(x^{s})))],
	\end{split}
	\end{equation}
	where the discriminator tries to maximize this objective to distinguish between $G_t(x^s)\! = \! x^{s\to t}$ and $x^t$, and meanwhile, the generator needs to minimize this objective to transform $x^s$ into realistic target-like images.

	To encourage the transformed images preserve contents of original images, a reverse generator is usually used to impose the cycle consistency~\cite{DBLP:conf/iccv/ZhuPIE17}.
	As shown in Fig.~\ref{fig:overview}, the feature encoder $E$ and upsampling decoder $U$ form the reverse target-to-source generator $G_s \! = \!E \circ U $ to reconstruct the $x^{s\to t}$ back to the source domain, and a discriminator $D_s$ operates in the source domain.
	This pair of source $\{G_s, D_s\}$ are trained in the same manner as $\{G_t, D_t\}$ with the adversarial loss $\mathcal{L}_{\textit{adv}}^s$.
	The reconstructed source images $x^{s\to t\to s}=U(E(G_t(x^s)))$ and the reconstructed target images $x^{t\to s \to t}=G_t(U(E(x^t)))$ are encouraged to be close to the original images $x^s$ and $x^t$ with the pixel-wise cycle-consistency loss:
	\begin{equation}
	\small
	\begin{split}
	\mathcal{L}_{\textit{cyc}}(G_t,E,U) = & ~ \mathbb{E}_{x^{s}\sim X^s}||U(E(G_t(x^s)))-x^s||_1+\\
	&~\mathbb{E}_{x^{t}\sim X^t}||G_t(U(E(x^t)))-x^t||_1.
	\end{split}
	\end{equation}
	
	\textit{2) Segmentation Network for Target Data:} ideally, this image-to-image transformation could bring $x^{s\to t}$ into the data distribution of target domain, such that these synthesized images can be used to train a segmentation network for the target domain.
	Specifically, after extracting features from the adapted image $x^{s\to t}$ with the encoder $E$, the feature maps $E(x^{s\to t})$ are forwarded to a pixel-wise classifier $C$ for predicting segmentation masks. In other words, the composition of $E \circ C$ serves as the segmentation network for the target domain.
	This part is trained with the sample pairs $\{x^{s\to t}, y^s\}$ by minimizing a hybrid loss $\mathcal{L}_{\textit{seg}}$ defined as:
	\begin{equation}
	\small
	\mathcal{L}_{\textit{seg}}(E,C)=H(y^s,C(E({x}^{s\to t}))+\textit{Dice}(y^s,C(E({x}^{s\to t}))),
	\end{equation}
	where the first term represents cross-entropy loss, the second term is the Dice loss.
	The hybrid loss function is designed to tackle the class imbalance in medical image segmentation~\cite{DBLP:conf/ijcai/DouOCCH18}.
	
	\subsection{Adversarial Learning for Feature Alignment}
	
	With above image alignment, training the segmentation network $E \circ C$ with the transformed target-like images can already get appealing performance on the target data.
	Unfortunately, when domain shift is severe, such as for cross-modality medical images, it is still insufficient to achieve the desired domain adaptation results. To this end, we impose additional discriminators to contribute from the perspective of feature alignment, to further reduce the domain gap between the synthesized target images $x^{s\to t}$ and real target images $x^t$.
	
	To align the extracted features of $x^{s\to t}$ and $x^t$, the most common way is applying adversarial learning directly in feature space, such that a discriminator fails to differentiate which features come from which domain. However, a feature space is with high-dimension, and hence difficult to be directly aligned. Instead, we choose to enhance the domain-invariance of feature distributions by using adversarial learning via two compact lower-dimensional spaces.
	Specifically, we inject adversarial losses via the semantic prediction space and the generated image space.
	
	\textit{1) Feature Alignment in Semantic Prediction Space:} as shown in Fig.~\ref{fig:overview}, for prediction of segmentation masks from $E \circ C$, we construct the discriminator $D_p$ to classify the outputs corresponding to $x^{s\to t}$ or $x^t$. 
	The semantic prediction space represents the information of human-body anatomical structures, which should be consistent across different imaging modalities. If the features extracted from $x^{s\to t}$ are aligned with that from $x^t$, the discriminator $D_p$ would fail in differentiating their corresponding segmentation masks, as the anatomical shapes are consistent. Otherwise, the adversarial gradients are back-propagated to the feature extractor $E$, so as to minimize the distance between the feature distributions from $x^{s\to t}$ and $x^t$. 
	The adversarial loss from semantic-level supervision for the feature alignment is: 
	\begin{equation}
	\small
	\begin{split}
	\mathcal{L}_{\textit{adv}}^{p}(E,C,D_p)= &\mathbb{E}_{x^{s\to t}\sim X^{s\to t}}[\text{log}~D_p(C(E(x^{s\to t})))]+\\
	&\mathbb{E}_{x^{t}\sim X^t}[\text{log}(1-D_p(C(E(x^{t}))))].
	\end{split}
	\end{equation}

	\textit{2) Deeply Supervised Adversarial Learning in Semantic Prediction Space:}
	as the adversarial gradients are back-propagated from the semantic prediction space to align feature distributions, the low-level features at lower layers, which are farther away from the compact space, may not be aligned as well as high-level features. In this regard, we introduce the deep supervision mechanism into the feature alignment to directly guide the training of both upper and lower layers so as to enhance the propagation of gradients flow to the low-level features. 
	Specifically, we connect an additional pixel-wise classifier with the outputs of lower layers of the encoder to make auxiliary predictions. Then a discriminator is constructed to differentiate those auxiliary predictions corresponding to $x^{s\to t}$ or $x^t$. This deeply supervised adversarial loss contributes to enhance the alignment of low-level features. In this regard, the segmentation loss in Equation (3) and the adversarial loss in Equation (4) can be extended as $\mathcal{L}_{\textit{seg}}^i(E,C_i)$ and $\mathcal{L}_{\textit{adv}}^{p_i}(E,C_i,D_{p_i})$ with $i=\{1,2\}$, where $C_1$ and $C_2$ denote the two classifiers connecting to different layers of the encoder, and $D_{p_1}$ and $D_{p_2}$ denote the two discriminators to differentiate the outputs of the two classifiers respectively.

	\textit{3) Feature Alignment in Generated Image Space:} for the generated source-like images from the target-to-source generator, that is the composition of $E \circ U$, we add an auxiliary task to the source discriminator $D_s$ to differentiate whether the generated images are reconstructed from $x^{s\to t}$ or transformed from real target images $x^t$. 
	If the discriminator $D_s$ succeeded in classifying the domain of generated images, it means that the extracted features still contain domain characteristics. To make the features domain-invariant, the following adversarial loss is employed to supervise the feature extraction process:
	\begin{equation}
	\small
	\begin{split}
	\mathcal{L}_{\textit{adv}}^{\tilde{s}}(E,D_s) = & ~\mathbb{E}_{x^{s\to t}\sim X^{s\to t}}[\text{log} D_s(U(E(x^{s\to t})))] + \\
	&~\mathbb{E}_{x^{t}\sim X^t}[\text{log}(1-D_s(U(E(x^{t}))))].
	\end{split}
	\end{equation}
	
	It is noted that the encoder $E$ is encouraged to extract features with domain-invariance by connecting discriminators from two aspects, i.e.,
	segmentation predictions and generated source-like images.
	By adversarial learning from these lower-dimensional compact spaces, the domain gap between synthesized target images $x^{s\to t}$ and real target images $x^t$ can be effectively addressed.
	
	\subsection{Shared Encoder for Synergistic Learning}
	
	Importantly, a key characteristic in our proposed synergistic learning framework is to share the feature encoder $E$ between both image and feature alignment.
	More specifically, encoder $E$ is optimized with the adversarial loss $\mathcal{L}_{\textit{adv}}^s$ and cycle-consistency loss $\mathcal{L}_{\textit{cyc}}$ via the image alignment process. It also collects gradients back-propagated from the discriminators $\{D_{p_i}, D_s\}$ towards feature alignment.
	In these regards, the feature encoder is fitted inside a multi-task learning scenario, such that, it is able to present generic and robust representations useful for multiple purposes.
	In turn, the different tasks bring complementary inductive bias to the encoder parameters, i.e., either emphasizing pixel-wise cyclic reconstruction or focusing on structural semantics.
	This can also contribute to alleviate the over-fitting problem with limited medical datasets when training such a complicated model.
	
	With the encoder enabling seamless integration of the image and feature alignment, we can train the unified framework in an end-to-end manner.
	At each training iteration, all the modules are sequentially updated in the following order: $G_t$ $\rightarrow$ $D_t$ $\rightarrow$ $E$ $\rightarrow$ $C_i$ $\rightarrow$ $U$ $\rightarrow$ $D_s$ $\rightarrow$ $D_{p_i}$. 
	Specifically, the generator $G_t$ is updated first to obtain the transformed target-like images. Then the discriminator $D_t$ is updated to differentiate the target-like images $x^{s\to t}$ and the real target images $x^t$. Next, the encoder $E$ is updated for feature extraction from $x^{s\to t}$ and $x^t$, followed by the updating of pixel-wise classifier $C_i$ and decoder $U$ to map the extracted features to the segmentation predictions and generated source-like images. Finally, the discriminator $D_s$ and $D_{p_i}$ are updated to classify the domain of their inputs to enhance feature-invariance.
	The overall objective for our framework is as follows:
	\begin{equation}
	\small
	\begin{split}
	\mathcal{L} =&~\mathcal{L}_{\textit{adv}}^{t}(G_t,D_t)+\lambda_{\textit{adv}}^{s}\mathcal{L}_{\textit{adv}}^{s}(E,U,D_s)+\\    &~\lambda_{\textit{cyc}}\mathcal{L}_{\textit{cyc}}(G_t,E,U)
	+\lambda_{\textit{seg}}^1\mathcal{L}_{\textit{seg}}^1(E,C_1)+\\
	&~\lambda_{\textit{seg}}^2\mathcal{L}_{\textit{seg}}^2(E,C_2)
	+\lambda_{\textit{adv}}^{p_1}\mathcal{L}_{\textit{adv}}^{p_1}(E,C,D_{p_1})+\\
	&~\lambda_{\textit{adv}}^{p_2}\mathcal{L}_{\textit{adv}}^{p_2}(E,C,D_{p_2})+\lambda_{\textit{adv}}^{\tilde{s}}\mathcal{L}_{\textit{adv}}^{\tilde{s}}(E,D_s),\\
	\end{split}
	\end{equation}
	where the $\{\lambda_{\textit{adv}}^{s},\lambda_{\textit{cyc}},\lambda_{\textit{seg}}^1,\lambda_{\textit{seg}}^2,\lambda_{\textit{adv}}^{p_1},\lambda_{\textit{adv}}^{p_2},\lambda_{\textit{adv}}^{\tilde{s}}\}$ are trade-off parameters adjusting the importance of each component, which are empirically set as $\{0.1, 10, 1.0, 0.1, 0.1, 0.01, 0.1\}$ respectively and are kept consistent for all the experiments.
	
	For testing on the images from the target domain, an image $x^t$ is forwarded into the encoder $E$, followed by applying the pixel-wise classifier $C_1$. In this way, the semantic segmentation result is obtained via $C_1(E(x^t))$, using the domain adaptation framework which is learned without the need of any target domain annotations.

	\subsection{Network Configurations and Implementation Details}
	Our network is composed of several 2D convolutional neural network modules. The layer configurations of the target generator $G_t$ follow the practice of CycleGAN \cite{DBLP:conf/iccv/ZhuPIE17}. It consists of 3 convolutional layers, 9 residual blocks, and 2 deconvolutional layers, followed by one convolutional layer to get the generated images. The source decoder $U$ is constructed with 1 convolutional layer, 4 residual blocks, and 3 deconvolutional layers, finally also followed by one convolutional output layer. 
	For all the discriminators $\{D_t, D_s, D_{p_i}\}$, the configurations follow PatchGAN \cite{DBLP:conf/cvpr/IsolaZZE17}, to differentiate $70 \! \times \! 70$ patches. The networks consist of 5 convolutional layers with kernels as size of $4 \! \times \! 4$ and stride of 2, except for the last two layers, which use convolution stride of 1. The numbers of feature maps are $\{64,128,256,512,1\}$ for each layer, respectively. At the first four layers, each convolutional layer is followed by an instance normalization and a leaky ReLU parameterized with $0.2$.
	
	The encoder $E$ uses residual connections and dilated convolutions ($\textit{dilation rate}\! = \! 2$) to enlarge the size of receptive field while preserving the spatial resolution for dense predictions~\cite{yu2017dilated}. 
	Let $\{\text{C}k,\text{R}k,\text{D}k\}$ denote a convolutional layer, a residual block and a dilated residual block with $k$ channels, respectively.
	The $\text{M}$ represents the max-pooling layer with a stride of 2.
	Our encoder module is deep by stacking layers of
	$\{\text{C16},\text{R16},\text{M},\text{R32},\text{M},2\!\times\!\text{R64},\text{M},2\!\times\!\text{R128},4\!\times\!\text{R256},2\!\times\!\text{R512},2\!\times\!\text{D512},2\!\times\!\text{C512}\}$. 
	Each convolution operation is connected to a batch normalization layer and ReLU activation. 
	The pixel-wise classifiers $C_1$ and $C_2$ is a $1\!\times\!1$ convolutional layer followed by an upsampling layer to recover the resolution of segmentation predictions to original image size. 
	The pixel-wise classifier $C_1$ is connected to the final outputs of the $2\!\times\!\text{C512}$ block, and the pixel-wise classifier $C_2$ is connected to the final outputs of the $2\!\times\!\text{R512}$ block of the encoder module respectively.
	
	We implemented our framework in TensorFlow (version 1.10.0). 
	Each model was trained 20k iterations with a batch size of 8 on one NVIDIA TITAN Xp GPU.
	In our earlier work~\cite{chen2019synergistic}, the adversarial learning and the segmentation task were optimized with different training strategies. 
	In this work, the whole network was optimized using the Adam optimizer with a learning rate of $2\!\times\!10^{-4}$. We found that the consistent learning rate for all network modules contribute to stabilize the model training process.

	\section{Experiments}
	
	We validated the effectiveness of our proposed unsupervised domain adaptation method with two applications on MRI and CT images, i.e., the cardiac substructure segmentation and the abdominal multi-organ segmentation. For both applications, the MRI and CT data are unpaired and collected from different patient cohorts. We conducted comprehensive evaluation for cross-modality adaptation in two directions, i.e., from MRI to CT images and CT to MRI images. 
	For each adaptation setting, the ground truth of target images were used for evaluation only, without being presented to the network during training phase. 
	
	\subsection{Dataset}
	\textbf{Cardiac substructure segmentation} We employed the Multi-Modality Whole Heart Segmentation (MMWHS) Challenge 2017 dataset for cardiac segmentation~\cite{zhuang2016multi}. 
	The training data consist of unpaired 20 MRI and 20 CT volumes with ground truth masks being provided. 
	We aim to adapt the segmentation network for parsing four cardiac structures including ascending aorta (AA), left atrium blood cavity (LAC), left ventricle blood cavity (LVC), and myocardium of the left ventricle (MYO).
	
	\textbf{Abdominal multi-organ segmentation} We utilized the T2-SPIR MRI training data from the ISBI 2019 CHAOS Challenge~\cite{CHAOS} with 20 volumes, and the public CT data from \cite{Synapse} with 30 volumes. The ground truth masks of four abdominal organs are provided in both datasets, including liver, right kidney, left kidney, and spleen, with which we conduct the task of multi-organ segmentation.
	
	For both applications, each modality was randomly split with 80\% scans for training and 20\% scans for testing. 
	The original MRI and CT scans have different field of view in both datasets. The cardiac MRI scans capture the area from neck to abdomen while the cardiac CT volumes consistently present the heart area. For the abdominal dataset, only the abdomen area is acquired in the MRI volumes while the area from neck to knee is captured in the CT volumes.
	In order to obtain similar field of view for all the volumes in each application, we manually cropped the original scans to cover the structures/organs which we aim to segment.
	For the cardiac dataset, we used a 3D bounding box with a fixed coronal plane size of $256\!\times\!256$ to crop a data volume, which centers on the heart area. For the abdominal dataset, we removed the axial slices that do not contain any of the four abdominal organs and cropped out the black background on the axial plane of each scan.
	All the data were normalized as zero mean and unit variance. 
	Each volume was re-sampled into the size of $256\!\times\!256$ and we employed data augmentation with rotation, scaling, and affine transformations to reduce over-fitting.
	To train our 2D networks for volumetric data segmentation, we used the coronal view image slices of the cardiac volumes and the axial view image slices of the abdominal scans.
	
	\subsection{Evaluation Metrics}
	We employed two commonly-used metrics, the Dice similarity coefficient (Dice) and the average symmetric surface distance (ASD), to quantitatively evaluate the segmentation performance of models. Dice measures the voxel-wise segmentation accuracy between the predicted and reference volumes. ASD calculates the average distances between the surface of the prediction mask and the ground truth in 3D. A higher Dice value and a lower ASD value indicate better segmentation results. 
	The evaluation is performed on the subject-level segmentation volume to be consistent with the MMWHS and CHAOS challenges as well as previous works~\cite{huo2018synseg,dou2018pnp}.
	
	\begin{table*}[t!]
		\caption{Performance comparison with different unsupervised domain adaptation methods for cardiac segmentation.
		}
		
		\centering
		\begin{center}
			\resizebox{0.75\textwidth}{!}{%
				\begin{tabular}{c|ccccc|ccccc}
					\toprule[1.0pt]
					
					\multicolumn{11}{c}{Cardiac MRI $\rightarrow$ Cardiac CT}\\	
					\hline
					\multirow{2}{*}{Method} &\multicolumn{5}{c|}{Dice}&\multicolumn{5}{c}{ASD}\\
					\cline{2-11}
					&AA &LAC &LVC &MYO &Average &AA &LAC &LVC &MYO &Average \\
					
					\hline
					
					Supervised training &92.7 &91.1 &91.9 &87.7 &90.9 &1.5 &3.5 &1.7 &2.1 &2.2 \\
					
					W/o adaptation &28.4 &27.7 &4.0 &8.7 &17.2 &20.6 &16.2 &N/A &48.4 &N/A\\
					
					\hline
					PnP-AdaNet\cite{dou2018pnp} &74.0 &68.9 &61.9 &50.8 &63.9 &12.8 &6.3 &17.4 &14.7 &12.8 \\
					
					SynSeg-Net~\cite{huo2018synseg} &71.6 &69.0 &51.6 &40.8 &58.2 &11.7 &7.8 &7.0 &9.2 &8.9 \\			
					
					AdaOutput~\cite{tsai2018learning} &65.2 &76.6 &54.4 &43.6 &59.9 &17.9 &\textbf{5.5} &5.9 &8.9 &9.6 \\
					
					CycleGAN~\cite{DBLP:conf/iccv/ZhuPIE17} &73.8 &75.7 &52.3 &28.7 &57.6 &11.5 &13.6 &9.2 &8.8 &10.8\\
					
					CyCADA~\cite{hoffman2017cycada} &72.9 &77.0 &62.4 &45.3 &64.4 &9.6 &8.0 &9.6 &10.5 &9.4 \\
					
					Prior SIFA~\cite{chen2019synergistic} &81.1 &76.4 &\textbf{75.7} &58.7 &73.0 &10.6 &7.4 &6.7 &\textbf{7.8} &8.1\\
					
					SIFA (Ours) &\textbf{81.3} &\textbf{79.5} &73.8 &\textbf{61.6} &\textbf{74.1} &\textbf{7.9} &6.2 &\textbf{5.5} &8.5 &\textbf{7.0}\\
					
					\bottomrule[1.0pt]
			\end{tabular}}
		\end{center}
		
		\centering
		\begin{center}
			\resizebox{0.75\textwidth}{!}{%
				\begin{tabular}{c|ccccc|ccccc}
					\toprule[1.0pt]
					
					\multicolumn{11}{c}{Cardiac CT $\rightarrow$ Cardiac MRI}\\	
					
					\hline
					
					\multirow{2}{*}{Method} &\multicolumn{5}{c|}{Dice}&\multicolumn{5}{c}{ASD}\\
					\cline{2-11}
					&AA &LAC &LVC &MYO &Average &AA &LAC &LVC &MYO &Average \\
					
					\hline
					
					Supervised training &82.8 &80.5 &92.4 &78.8 &83.6 &3.6 &3.9 &2.1 &1.9 &2.9\\
					
					W/o adaptation &5.4 &30.2 &24.6 &2.7 &15.7 &15.4 &16.8 &13.0 &10.8 &14.0\\
					
					\hline
					PnP-AdaNet~\cite{dou2018pnp} &43.7 &47.0 &77.7 &\textbf{48.6} &54.3 &11.4 &14.5 &4.5 &5.3 &8.9 \\
					
					SynSeg-Net~\cite{huo2018synseg} &41.3 &57.5 &63.6 &36.5 &49.7 &8.6 &10.7 &5.4 &5.9 &7.6 \\
					
					AdaOutput~\cite{tsai2018learning} &60.8 &39.8 &71.5 &35.5 &51.9 &\textbf{5.7} &8.0 &4.6 &4.6 &\textbf{5.7} \\
					
					CycleGAN~\cite{DBLP:conf/iccv/ZhuPIE17} &64.3 &30.7 &65.0 &43.0 &50.7 &5.8 &9.8 &6.0 &5.0 &6.6\\		
					
					CyCADA~\cite{hoffman2017cycada} &60.5 &44.0 &77.6 &47.9 &57.5 &7.7 &13.9 &4.8 &5.2 &7.9 \\
					
					Prior SIFA~\cite{chen2019synergistic} &\textbf{67.0} &60.7 &75.1 &45.8 &62.1 &6.2 &9.8 &4.4 &\textbf{4.4} &6.2\\
					
					SIFA (Ours) &65.3 &\textbf{62.3} &\textbf{78.9} &47.3 &\textbf{63.4} &7.3 &\textbf{7.4} &\textbf{3.8} &\textbf{4.4} &\textbf{5.7}\\
					
					\bottomrule[1.0pt]
			\end{tabular}}
		\end{center}
		\vspace{-0mm}
		\label{tab:cardiac}
	\end{table*}

	\begin{figure*}[h!]
		\centering
		\includegraphics[width=1\textwidth]{{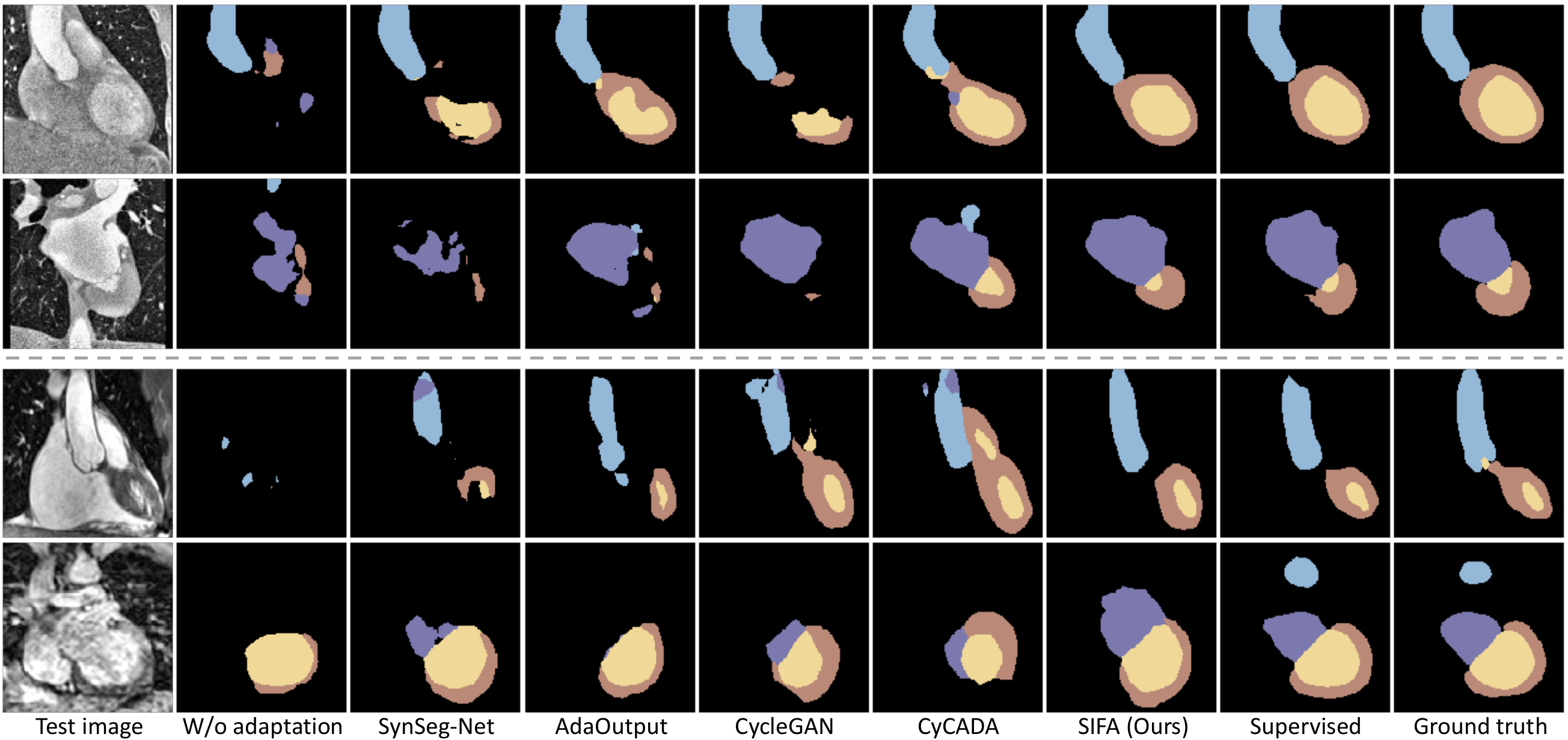}}
		\caption{Visual comparison of segmentation results produced by different methods for cardiac CT images (top two rows) and MRI images (bottom two rows). From left to right are the raw test images (1st column), ``W/o Adaptation" lower bound (2nd column), results of other unsupervised domain adaptation methods (3rd-6th column), results of our SIFA network (7th column), results of supervised training (8th column), and ground truth (last column). The cardiac structures of AA, LAC, LVC, and MYO are indicated in blue, purple, yellow, and brown color respectively. Each row corresponds to one example.
		}
		\label{fig:cardiac}
		\vspace{-0mm}
	\end{figure*}

	\subsection{Effectiveness of SIFA on Unsupervised Domain Adaptation}
	To observe the effect of domain shift on segmentation performance, we first obtain the ``W/o adaptation" lower bound by directly applying the model learned in source domain to test target images without using any domain adaptation method. We also provide the performance upper bound of supervised training with target domain labels to measure the performance gap. The segmentation results produced by our method are then compared with the lower and upper bounds to validate the effectiveness of our method on reducing the severe performance degradation caused by domain shift. For a consistent comparison, the segmentation network architecture in the SIFA framework is adopted for training the lower and upper bounds models, that is the composition of the encoder $E$ and the pixel-wise classifier $C_1$.
	
	Table~\ref{tab:cardiac} reports the segmentation results for cardiac datasets. Without domain adaptation, the model trained on MRI images only obtained the average Dice of 17.2\% when being tested on CT images directly, and the average Dice of 15.7\% for the reverse direction. The significant performance gap to the supervised training upper bound is 73.7 percentage points for CT images and 67.9 percentage points for MRI images. 
	This demonstrates the severe domain shift between MRI and CT images, which led to similar level of performance degradation on cross-modality segmentation when employing either type of modality as the source domain.  
	Remarkably, our SIFA network consistently improves the cross-modality segmentation performance to a large degree in terms of both Dice and ASD measurements. 
	For CT images, we improved the average Dice to 74.1\% over the four cardiac structures with the average ASD being reduced to 7.0, and for MRI images, we achieved the average Dice of 63.4\% and the average ASD 5.7. 
	The qualitative segmentation results in Fig.~\ref{fig:cardiac} also show that without adaptation, it is difficult to obtain correct prediction for any cardiac structure. Instead, our method can successfully locate the four cardiac structures and generate semantically meaningful segmentation. 
	Both the quantitative and qualitative results validate the effectiveness of our method on addressing the severe domain shift.
	
	\begin{table*}[h!]
		\caption{Performance comparison with different unsupervised domain adaptation methods for abdominal segmentation.
		}
		\centering
		\begin{center}
			\resizebox{0.95\textwidth}{!}{%
				\begin{tabular}{c|ccccc|ccccc}
					\toprule[1.0pt]
					
					\multicolumn{11}{c}{Abdominal MRI $\rightarrow$ Abdominal CT}\\	
					\hline
					\multirow{2}{*}{Method} &\multicolumn{5}{c|}{Dice}&\multicolumn{5}{c}{ASD}\\
					\cline{2-11}
					&Liver &R. kidney &L. kidney &Spleen &Average &Liver &R. kidney &L. kidney &Spleen &Average \\
					
					\hline
					
					Supervised training &92.8 &86.4 &87.4 &88.2 &88.7 &1.0 &1.8 &0.9 &1.2 &1.2\\
					
					W/o adaptation &73.1 &47.3 &57.3 &55.1 &58.2 &2.9 &5.6 &7.7 &7.4 &5.9\\
					
					\hline
					
					SynSeg-Net~\cite{huo2018synseg} &85.0 &82.1 &72.7 &81.0 &80.2 &2.2 &1.3 &2.1 &2.0 &1.9 \\		
					
					AdaOutput~\cite{tsai2018learning} &85.4 &79.7 &79.7 &81.7 &81.6 &1.7 &1.2 &1.8 &\textbf{1.6} &1.6 \\
					
					CycleGAN~\cite{DBLP:conf/iccv/ZhuPIE17} &83.4 &79.3 &79.4 &77.3 &79.9 &1.8 &1.3 &\textbf{1.2} &1.9 &1.6\\
					
					CyCADA~\cite{hoffman2017cycada} &84.5 &78.6 &80.3 &76.9 &80.1 &2.6 &1.4 &1.3 &1.9 &1.8 \\
					
					Prior SIFA~\cite{chen2019synergistic} &87.9 &\textbf{83.7} &80.1 &80.5 &83.1 &2.1 &1.1 &1.6 &1.8 &1.6\\
					
					SIFA (Ours) &\textbf{88.0} &83.3 &\textbf{80.9} &\textbf{82.6} &\textbf{83.7} &\textbf{1.2} &\textbf{1.0} &1.5 &\textbf{1.6} &\textbf{1.3}\\
					
					\bottomrule[1.0pt]
			\end{tabular}}
		\end{center}
		
		\centering
		\begin{center}
			\resizebox{0.95\textwidth}{!}{%
				\begin{tabular}{c|ccccc|ccccc}
					\toprule[1.0pt]
					
					\multicolumn{11}{c}{Abdominal CT $\rightarrow$ Abdominal MRI}\\	
					
					\hline
					
					\multirow{2}{*}{Method} &\multicolumn{5}{c|}{Dice}&\multicolumn{5}{c}{ASD}\\
					\cline{2-11}
					&Liver &R. kidney &L. kidney &Spleen &Average &Liver &R. kidney &L. kidney &Spleen &Average \\
					
					\hline
					
					Supervised training &92.0 &91.1 &80.6 &85.7 &87.3 &1.3 &2.0 &1.5 &1.3 &1.5\\
					
					W/o adaptation &48.9 &50.9 &65.3 &65.7 &57.7 &4.5 &12.3 &6.8 &4.5 &7.0\\

					\hline
					SynSeg-Net~\cite{huo2018synseg} &87.2 &\textbf{90.2} &76.6 &79.6 &83.4 &2.8 &0.7 &4.8 &2.5 &2.7 \\
					
					AdaOutput~\cite{tsai2018learning} &85.8 &89.7 &76.3 &82.2 &83.5 &1.9 &1.4 &3.0 &1.8 &2.1 \\
					
					CycleGAN~\cite{DBLP:conf/iccv/ZhuPIE17} &88.8 &87.3 &76.8 &79.4 &83.1 &2.0 &3.2 &1.9 &2.6 &2.4\\
					
					CyCADA~\cite{hoffman2017cycada} &88.7 &89.3 &78.1 &80.2 &84.1 &\textbf{1.5} &1.7 &\textbf{1.3} &\textbf{1.6} &\textbf{1.5} \\
					
					Prior SIFA~\cite{chen2019synergistic} &88.5 &90.0 &79.7 &81.3 &84.9 &2.3 &0.9 &1.4 &2.4 &1.7\\

					SIFA (Ours) &\textbf{90.0} &89.1 &\textbf{80.2} &\textbf{82.3} &\textbf{85.4} &\textbf{1.5} &\textbf{0.6} &1.5 &2.4 &\textbf{1.5}\\

					\bottomrule[1.0pt]
			\end{tabular}}
		\end{center}
		\label{tab:organ}
	\end{table*}
	\begin{figure*}[h!]
		\centering
		\includegraphics[width=1\textwidth]{{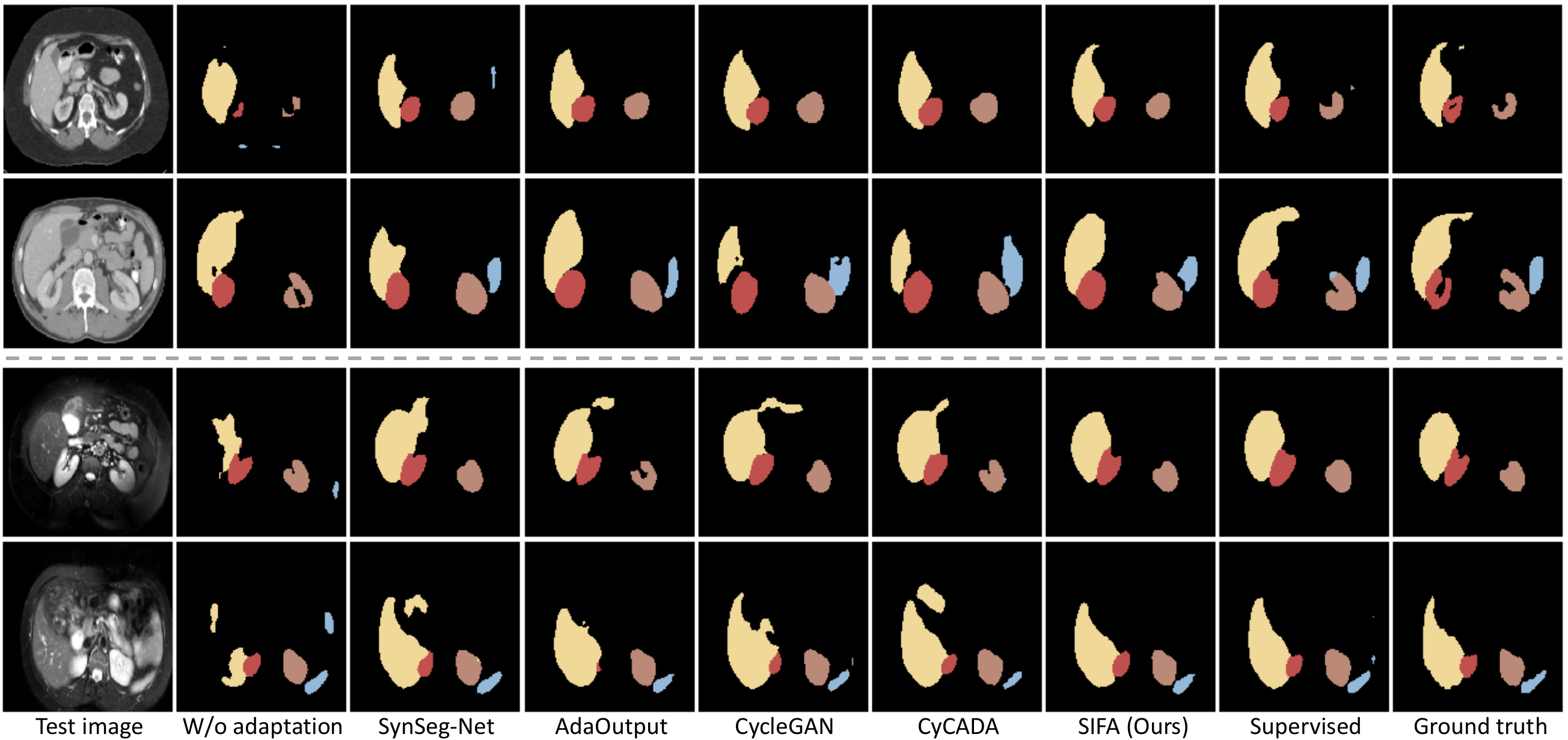}}
		\caption{Visual comparison of segmentation results produced by different methods for abdominal CT images (top two rows) and MRI images (bottom two rows). From left to right are the raw test images (1st column), ``W/o Adaptation" lower bound (2nd column), results of other unsupervised domain adaptation methods (3rd-6th column), results of our SIFA network (7th column), results of supervised training (8th column), and ground truth (last column). The liver, right kidney, left kidney, and spleen are indicated in yellow, red, brown, and blue color respectively. Each row corresponds to one example.
		}
		\label{fig:organ}
	\end{figure*}
	Table~\ref{tab:organ} presents the results for segmentation in abdominal images, where we can observe cross-modality performance degradation similar to the cardiac dataset. Without domain adaptation, the average Dice is only 58.2\% for segmenting abdominal CT images, and 57.7\% for MRI images. It is noted that for both domains there exists about 30 percentage points performance gap to the supervised training upper bound, which is not as severe as the cardiac dataset. It maybe because it is harder to locate the correct cardiac structure than abdominal organs and the variation of image quality in cardiac data is higher than the abdominal data. For example, as shown in the last row of Fig.~\ref{fig:cardiac}, it is very difficult to identify the LAC structure in the cardiac MRI images because of its limited contrast with the surrounding tissue. While for abdominal images as we can see in the second column of Fig.~\ref{fig:organ}, predicting the approximate location of organs is easier than the cardiac structures, though delineating the accurate boundary is also challenging. Remarkably, our SIFA achieved the average Dice of 83.7\% and average ASD of 1.3 for CT images, and average Dice of 85.4\% and average ASD of 1.5 for MRI images, which are very close to the supervised training upper bound. For CT images, the gap is 5 percentage points in Dice, 0.1 in ASD, and for MRI images, the gap is only 1.9 percentage points in Dice with the same ASD. This demonstrates that unsupervised domain adaptation is very promising and has the potential to have practical values. 
	Fig.~\ref{fig:organ} shows that our SIFA can successfully perform segmentation with satisfactory boundary delineation on the four different organs which have significant variations in size and shape. 
	
	\subsection{Comparison with State-of-the-art Methods}
	We compare our proposed method with recent state-of-the-art unsupervised domain adaptation approaches including PnP-AdaNet~\cite{dou2018pnp}, SynSeg-Net~\cite{huo2018synseg}, AdaOutput~\cite{tsai2018learning}, CycleGAN~\cite{DBLP:conf/iccv/ZhuPIE17}, and CyCADA~\cite{hoffman2017cycada}, which utilize either feature alignment, image alignment or their mixtures. 
	The first two are dedicated to cross-modality segmentation on MRI/CT images using feature alignment and image alignment respectively. 
	The last three are well-established methods on natural image datasets. AdaOutput employs feature alignment, CycleGAN adapts image appearance, and CyCADA conducts both image and feature alignments. 
	For a fair comparison, the segmentation network architecture used for the implementation of other methods is the same as that used in SIFA. For PnP-AdaNet on cardiac dataset, we directly referenced the results from their paper as they used the same dataset and segmentation backbone as ours.
	
	The quantitative performance of different methods are presented in Table~\ref{tab:cardiac} and Table~\ref{tab:organ}, and the visual comparison results are shown in Fig.~\ref{fig:cardiac} and Fig.~\ref{fig:organ} for cardiac images and abdominal images respectively. It is observed that our method significantly outperforms other comparison approaches by a large margin.
	Notably, CyCADA and our SIFA, which utilize both image and feature alignments, obtained better segmentation results than other methods which adopt either one aspect of alignment. This shows that the two adaptive processes can be conducted altogether to achieve a stronger domain adaptation model. 
	Compared with CyCADA, our method achieved superior performance, demonstrating the effectiveness of our synergistic learning framework in unleashing the mutual benefits of image and feature alignments. 
	It is worth noting that, the performance of CyCADA on segmenting the abdominal CT data is worse than the feature alignment method AdaOutput, indicating that how to combine the two types of alignment to effectively utilize their advantages is of importance. With our synergistic learning strategy, the segmentation results can be further boosted with higher increase than using image alignment or feature alignment alone. 
	
	\subsection{Ablation Analysis of Key Components}
	\begin{figure}[t]
		\includegraphics[width=0.45\textwidth]{{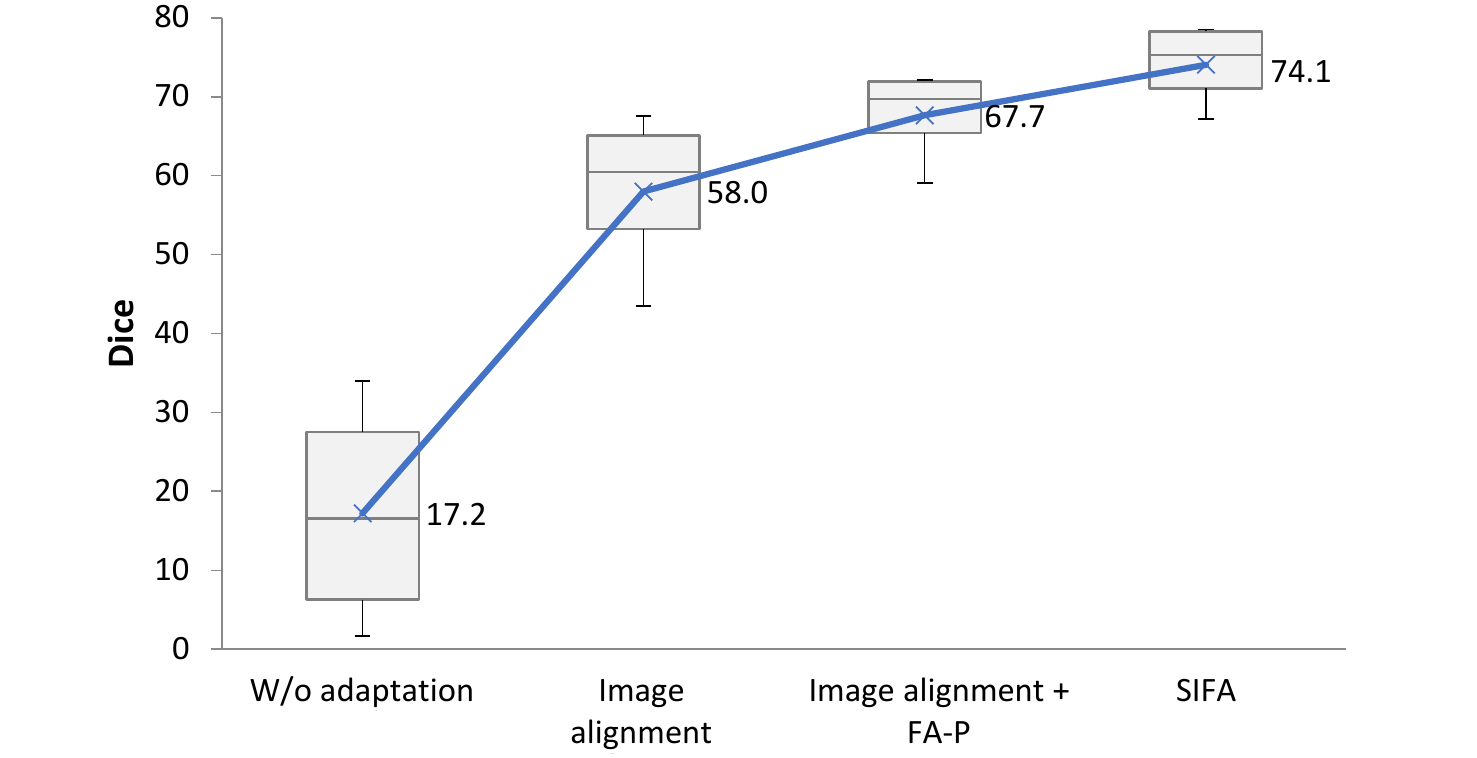}}
		\caption{Boxplot of segmentation results in each ablation experimental setting for analysis of the proposed SIFA framework. The confidence intervals are generated based on the different volumes in the test sets.
		}
		\label{fig:boxplot}
	\end{figure}
	We first conduct ablation experiments on the cardiac dataset for MRI to CT adaptation to demonstrate the image alignment and the feature alignment in two compact spaces can work jointly to improve domain adaptation performance. 
	The results are presented in Fig.~\ref{fig:boxplot}.
	Our baseline network uses image alignment only, which is constructed by removing the feature alignment adversarial loss $\{\mathcal{L}_{\textit{adv}}^{p_1},\mathcal{L}_{\textit{adv}}^{p_2},\mathcal{L}_{\textit{adv}}^{\tilde{s}}\}$ when training the network. Compared with the ``W/o adaptation" lower bound, our baseline network with image alignment alone increased the average Dice to 58.0\%. This shows that with image transformation, the source images have been brought closer to the target domain successfully. 
	Then we add the deeply supervised feature alignment in the semantic prediction space, i.e., FA-P\textsubscript{1} and FA-P\textsubscript{2}, into the baseline network, which increased the average Dice to 67.7\%.
	Further adding the feature alignment in the generated image space, i.e., FA-I, corresponds to our proposed SIFA model, which obtained average Dice of 74.1\%.
	The continuous increase in segmentation accuracy demonstrates that the image and feature alignment can be jointly conducted to achieve better domain adaptation, and the feature alignment in different compact spaces could inject effects from integral aspects to encourage domain invariance.
	
	\begin{figure}[t]
		\includegraphics[width=0.48\textwidth]{{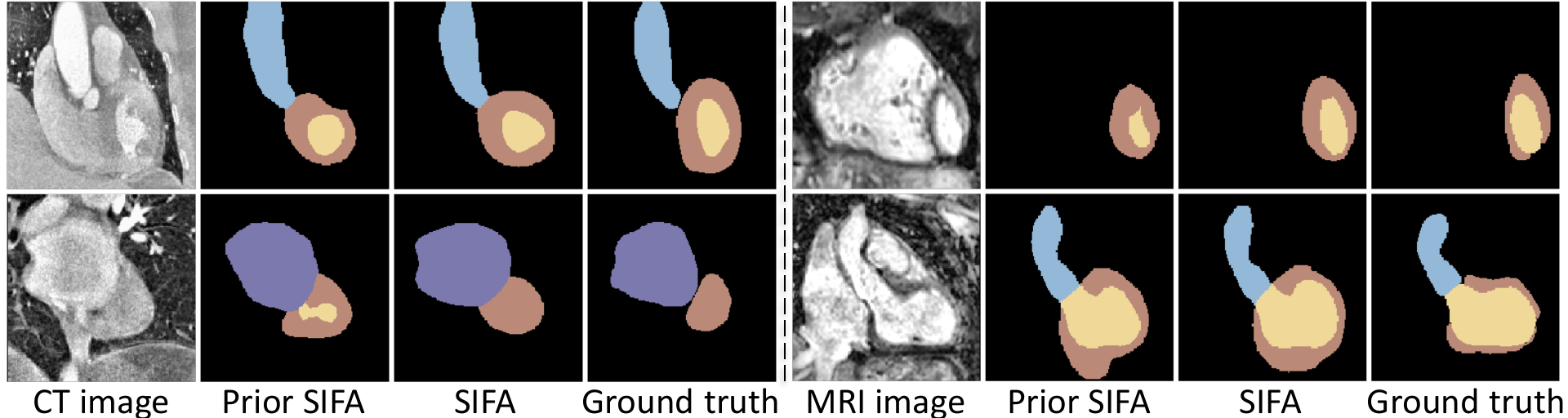}}
		\caption{Visual comparison of segmentation results of prior SIFA and SIFA for cardiac CT images (left) and MRI images (right). The cardiac structures of AA, LAC, LVC, and MYO are indicated in blue, purple, yellow, and brown color respectively.
		}
		\label{fig:ablation}
	\end{figure}
	
	\begin{figure}[t]
		\includegraphics[width=0.49\textwidth]{{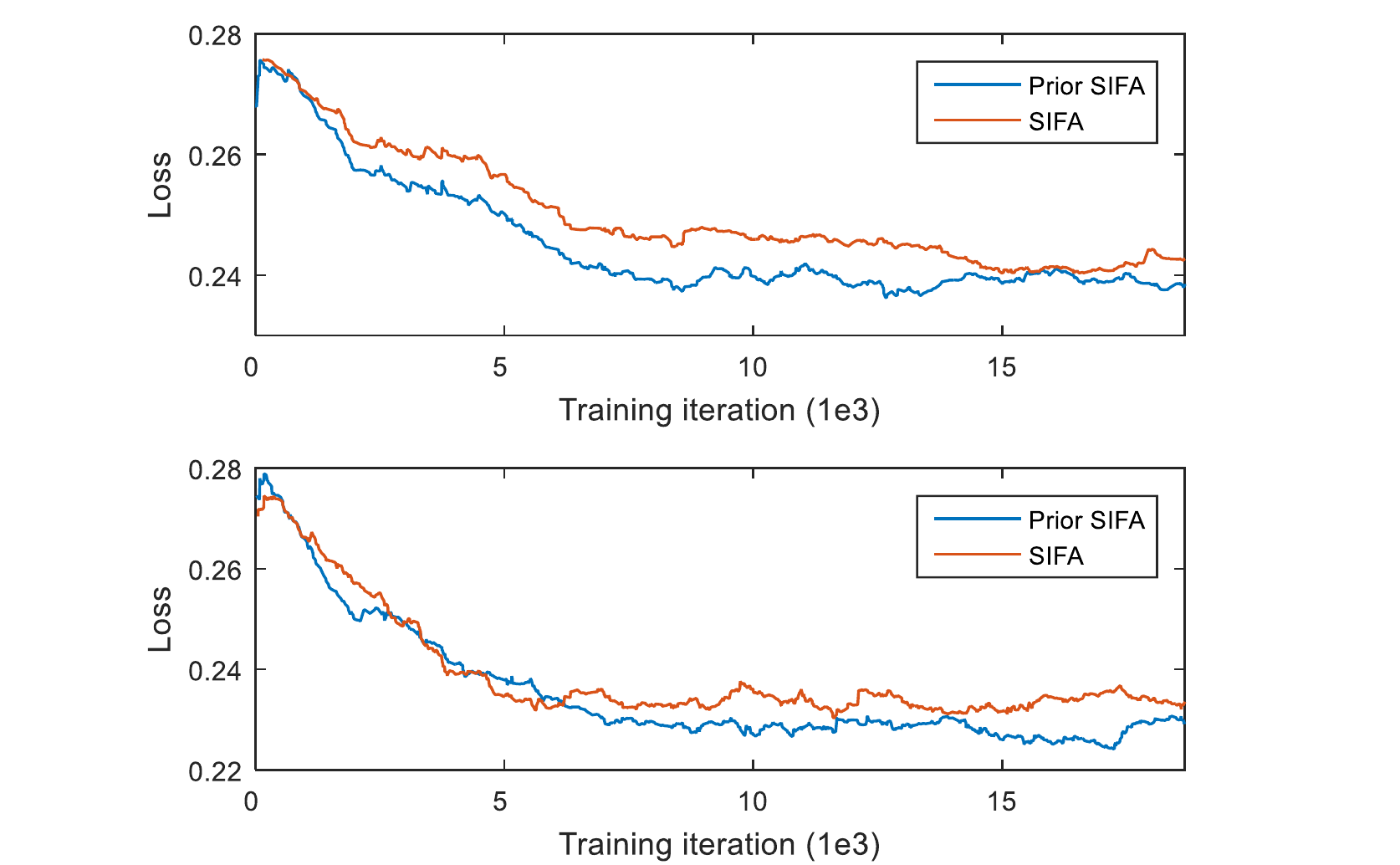}}
		\caption{Comparison of training loss curves of the discriminator $D_{p_1}$ between prior SIFA and SIFA. The top and bottom figures are from the cardiac MRI to CT adaptation and the cardiac CT to MRI adaptation, respectively.
		}
		\label{fig:losscurve}
	\end{figure}
	Then we further evaluate the deeply supervised mechanism introduced in this work by comparing to our prior conference paper~\cite{chen2019synergistic}, i.e., prior SIFA. The quantitative results of \cite{chen2019synergistic} are included in Table~\ref{tab:cardiac} and Table~\ref{tab:organ} for comparison.
	We can see that our proposed method consistently improves the domain adaptation performance over prior SIFA with higher average Dice and decreased ASD value, especially for structures with small size or irregular boundary, such as the cardiac LAC and MYO structures and the spleen in abdominal dataset. This indicates that the auxiliary adversarial learning in semantic prediction space, i.e., FA-P\textsubscript{2}, contributes to more effective alignment of low-level features so as to achieve better domain adaptation performance. 
	Fig.~\ref{fig:ablation} shows the visual comparison results on the cardiac dataset. As illustrated in the figure, the segmentation results produced by our proposed SIFA model match better with the ground truth, such as the shape and boundary of the MYO structure. In Fig.~\ref{fig:losscurve}, we show the training loss curves of the discriminator $D_{p_1}$ on cardiac dataset, which is connected to the outputs of the pixel-wise classifier $C_1$ and aims to differentiate the segmentation predictions of the transformed source images and the target images. The higher discriminator loss of our SIFA model than \cite{chen2019synergistic} indicates that the feature distributions have been aligned better such that it becomes harder for the discriminator to classify the domain of the segmentation predictions.
	
	\section{Discussion}
	In clinical practice, multiple imaging modalities are commonly used for measuring the same anatomy regarding to the complementary characteristics of different modalities~\cite{cao2017dual}. 
	Accurate segmentation of multi-modality images can be achieved with deep learning techniques, but labeled data is required for each modality because it is difficult for deep models to generalize well across modalities. 
	To alleviate the burden of data annotation, some works focus on cross-modality image synthesis so that the segmentation of multiple modalities can be achieved with synthesized images and one-modality labels~\cite{frangi2018simulation,hiasa2018cross,chartsias2017adversarial}.
	Recently, some works have explored the feasibility of cross-modality unsupervised domain adaptation to adapt deep models from the label-rich source modality to unlabeled target modality~\cite{DBLP:conf/ijcai/DouOCCH18,huo2018synseg}, with good results reported. 
	Our work proceeds along this promising direction, by demonstrating that without extra annotation effort, unsupervised domain adaptation can greatly reduce the performance degradation and for some tasks can even achieve very close segmentation performance to supervised training.
	
	The general efficacy of our method has been sufficiently demonstrated on two different multi-class segmentation tasks without specific network architecture or hyperparameter tuning.
	It is observed that the segmentation network has severe cross-modality performance degradation on both the cardiac datasets and the abdominal datasets. 
	With our proposed method, the segmentation performance for the target modality has been consistently and significantly improved in both tasks.
	However, the performance gap to the supervised training upper bound is larger in the cardiac application than the abdominal dataset. 
	We think the reasons could be two-fold. First, the task of cardiac segmentation itself maybe harder than the abdominal organ segmentation. In Fig.~\ref{fig:cardiac}, the LVC and MYO structures in cardiac CT images have very low contrast to the surrounding tissues, and it is very difficult to recognize the LAC structure in cardiac MRI images.
	Second, the cardiac dataset has various imaging quality with relatively poor quality for some data. For example, in Fig.~\ref{fig:cardiac}, the image resolution of the MRI test image in the last row is much lower than the one in the third row. 
	We argue that the task itself and varied data quality in the cardiac dataset increase the difficulty of domain adaptation. 
	For the abdominal images, the segmentation results of our method are very close to the supervised training upper bound in terms of both Dice and ASD values. This indicates that unsupervised domain adaptation is very promising and has the potential to have practical applications.
	
	For cross-modality adaptation, an important question is whether the adaptation is symmetric to modality, i.e., whether both the adaptations from MRI to CT and CT to MRI are feasible and whether the adaptation difficulty depends on the adaptation directions.
	To investigate that, we conduct bidirectional domain adaptation between MRI and CT images on both datasets, which has not been consistently conducted in current cross-modality works~\cite{DBLP:conf/ijcai/DouOCCH18,jiang2018tumor,huo2018synseg}.
	Our method greatly improves the segmentation performance for both adaptation directions of MRI to CT and CT to MRI, demonstrating that the cross-modality adaptation can be achieved in both directions. Meanwhile, it is observed that, in the cardiac segmentation, higher Dice can be achieved for adapting from MRI to CT domain than the other way around.
	Comparing the segmentation performance of supervised training on cardiac CT and MRI images (90.9\% versus 83.6\% in Dice), it is more difficult to segment the structures of interests in MRI images. This may make the adaptation from CT to MRI harder than the opposite direction.
	However, for the abdominal images, the adaptation performance for both directions are equivalently high and close to the supervised training upper bound.
	This indicates that the difficulty of domain adaptation across modalities might depend more on the task than the adaptation direction, 
	which adds new findings over the previous work~\cite{dou2018pnp}.
	Potential future studies in terms of different segmentation tasks will help further analyze this issue.
	
	One limitation of our work is using 2D networks for the volumetric image segmentation tasks.
	As our domain adaptation framework consists of multiple network modules and various aspects of adversarial learning, implementing such a complicated framework in 3D is memory intensive and training prohibitive.
	Existing works have demonstrated the feasibility of either unpaired 3D image synthesis~\cite{zhang2018translating} or the feature alignment with 3D networks~\cite{kamnitsas2017unsupervised}. However, achieving both the image and feature alignments in 3D model has not been explored yet, which lies in our future work.
	Another limitation comes from the relatively balanced amount of data in the source and target domains.
	In real-world clinical practice, there are two typical scenarios: 1) the available data in each domain can be imbalanced, e.g., the number of available CT scans is usually larger than MRI scans; 2) the number of target domain images may be limited. 
	Hence, in the future, we would like to explore effective unsupervised domain adaptation on imbalanced dataset as well as when only limited number of target domain data is available.
	
	Our current network architectures follow the practice of CycleGAN~\cite{DBLP:conf/iccv/ZhuPIE17} by using the Resnet blocks for the generator and decoder, and follow the previous cross-modality adaptation work~\cite{DBLP:conf/ijcai/DouOCCH18} for the configurations of the segmentation model. To validate the effectiveness of our segmentation backbone, we compare the supervised training performance of our segmentation model on cardiac dataset with Payer et al.~\cite{payer2017multi}, which obtained the first ranking in the MMWHS Challenge 2017. Table~\ref{tab:supervise} shows that our model can achieve comparable performance to Payer et al.
	But unlike ~\cite{DBLP:conf/ijcai/DouOCCH18,dou2018pnp} which prefers network architectures without skip connections, we consider that other network architectures could also be used in our framework, such as the Unet model~\cite{ronneberger2015u}, which is the most common networks for medical image segmentation. The generator can be directly implemented as a Unet model, which has been demonstrated to be effective and stable during training for unpaired image-to-image transformation\footnote{~\url{https://github.com/junyanz/pytorch-CycleGAN-and-pix2pix}}. Then the shared encoder in our framework can be configured as the downsampling part of a Unet model, and the decoder and pixel-wise classifier can both connect to the bottleneck layer and be configured as two separate upsampling parts of a Unet model.
	Although we argue that there is flexibility of designing network architectures, it is worth noting that the training stability and memory consumption are the two main considerations for network implementation in our framework.
	
	When deploying deep neural networks to medical images, the issue of domain shift is inevitable and widely exists. The large performance degradation of deep models has been observed between different MRI sequences in \cite{kamnitsas2017unsupervised}. Even when using the same MRI sequence, the data distribution can vary in datasets acquired at different centers~\cite{perone2019unsupervised,gibson2018inter} or different time~\cite{ghafoorian2017transfer}. 
	Besides MRI images, domain adaptation studies have been performed on cross-site ultrasound datasets~\cite{yang2018generalizing,degel2018domain}, X-ray images~\cite{dong2018unsupervised,chen2018semantic}, histopathology applications~\cite{ren2018adversarial}, and optical fundus imaging~\cite{wang2019patch}.
	Compared to above scenarios, cross-modality adaptation is perhaps the most challenging situation due to the significant domain shift caused by the different physical principles of modalities~\cite{DBLP:conf/ijcai/DouOCCH18,jiang2018tumor,huo2018synseg,joyce2018deep}.
	We validate our method with this challenging setting of adaptation between CT and MRI data. Our aim is to align the image and feature distributions between the two modalities, not assuming that the underlying physical mapping across modalities could be directly learned by the networks.
	Notably, our method is general and can be easily applied to improve the adaptation performance for other situations. 
	
	\begin{table}[t]
		\caption{Comparison of segmentation results in Dice for cardiac substructure segmentation.
		}
		\centering
		\begin{center}
			\resizebox{0.4\textwidth}{!}{%
				\begin{tabular}{c|ccccc}
					\toprule[1.0pt]
					
					\multicolumn{6}{c}{Cardiac CT}\\	
					\hline
					\multirow{2}{*}{Method} &\multicolumn{5}{c}{Dice}\\
					\cline{2-6}
					&AA &LAC &LVC &MYO &Average\\
					
					\hline
					
					Payer et al.~\cite{payer2017multi} &91.1 &92.4 &92.4 &87.2 &90.8\\
					Supervised training	&92.7 &91.1 &91.9 &87.7 &90.9\\
					
					\bottomrule[1.0pt]
			\end{tabular}}
		\end{center}
		
		\centering
		\begin{center}
			\resizebox{0.4\textwidth}{!}{%
				\begin{tabular}{c|ccccc}
					\toprule[1.0pt]
					
					\multicolumn{6}{c}{Cardiac MRI}\\	
					
					\hline
					
					\multirow{2}{*}{Method} &\multicolumn{5}{c}{Dice}\\
					\cline{2-6}
					&AA &LAC &LVC &MYO &Average\\
					
					\hline
					
					Payer et al.~\cite{payer2017multi} &76.6 &81.1 &87.7 &75.2 &80.2\\
					Supervised training	&82.8 &80.5 &92.4 &78.8 &83.6\\
					
					\bottomrule[1.0pt]
			\end{tabular}}
		\end{center}
		\label{tab:supervise}
	\end{table}
	
	\section{Conclusion}
	We propose a novel framework SIFA for unsupervised domain adaptation of medical image segmentation, which synergistically combines the image alignment and feature alignment. 
	The two adaptive perspectives are guided by the adversarial learning with a shared feature encoder to exploit their mutual benefits for reducing domain shift during the end-to-end training. 
	We validate our method for unpaired bidirectional adaptation on two challenging multi-class segmentation tasks.
	Experimental results demonstrate the effectiveness of our SIFA framework in improving the performance of segmentation network in the target domain.

	\bibliographystyle{IEEEtran.bst}
	\bibliography{reference}
	
\end{document}